\documentclass[english,aps,prd,preprint, superscriptaddress,nofootinbib,12pt]{revtex4-1}

\usepackage{graphicx} 

\usepackage[T1]{fontenc}
\usepackage{lmodern}
\usepackage{caption}
\usepackage{floatrow}

\usepackage{subcaption}
\usepackage{hyperref}
\hypersetup{colorlinks,linkcolor={blue},citecolor={blue},urlcolor={blue}}

\usepackage{xcolor}
\usepackage{amsmath}
\usepackage{amssymb}
\usepackage{slashed}

\usepackage{amsmath,amssymb,amsfonts}
\usepackage{amsmath}
\usepackage{graphicx}
\usepackage{comment}
\usepackage{mathtools}
\usepackage{slashed}
\usepackage{physics}
\usepackage{mathrsfs}
\usepackage{enumerate}

\usepackage{bm}
\usepackage[utf8]{inputenc}

\usepackage{listings}

\usepackage[T1]{fontenc}
\usepackage{lmodern}

\usepackage{hyperref}

\usepackage{dsfont}

\hypersetup{colorlinks,linkcolor={blue},citecolor={blue},urlcolor={blue}} 
\usepackage{braket}
\usepackage[english]{babel}

\usepackage{floatrow}

\usepackage[normalem]{ulem}
\usepackage{color}

\usepackage{array}
\def\beq{\begin{equation}}
\def\eeq{\end{equation}}
\def\be{\begin{equation}}
\def\ee{\end{equation}}
\def\bea{\begin{eqnarray}}
\def\eea{\end{eqnarray}}

\begin{document}

\title{Neutrino Lorentz invariance violation from the $CPT-$even SME coefficients through a tensor interaction with cosmological scalar fields}

\author{Rub\'en Cordero}\email{rcorderoe@ipn.mx} \affiliation{Departamento de F\'{\i}sica, Escuela Superior de F\'{\i}sica y Matem\'aticas del Instituto Polit\'ecnico Nacional, Unidad Adolfo L\'opez Mateos, Edificio 9, 07738 Ciudad de M\'exico, Mexico}
\author{Luis A. Delgadillo} \email{ldelgadillof2100@alumno.ipn.mx}
\affiliation{Departamento de F\'{\i}sica, Escuela Superior de F\'{\i}sica y Matem\'aticas del Instituto Polit\'ecnico Nacional, Unidad Adolfo L\'opez Mateos, Edificio 9, 07738 Ciudad de M\'exico, Mexico}
\affiliation{Institute of High Energy Physics, Chinese Academy of Sciences, Beijing 100049, China} 
\author{O.~G.~Miranda} \email{omar.miranda@cinvestav.mx}
\affiliation{Departamento de F\'{\i}sica, Centro de Investigaci{\'o}n y de Estudios Avanzados del IPN Apdo. Postal 14-740 07000 Ciudad de M\'exico, Mexico}
\author{C.~A.~Moura}%
\email{celio.moura@ufabc.edu.br}
\affiliation{Centro de Ci\^encias Naturais e Humanas, Universidade Federal do ABC - UFABC, Av. dos Estados, 5001, 09210-580, Santo Andr\'e-SP, Brazil}

\date{\today}

\begin{abstract}

\noindent
Numerous non-standard interactions between neutrinos and scalar fields have been suggested in the literature.
In this work, we have outlined the case of tensorial neutrino non-standard interactions with scalar fields, which can be related to the effective $CPT-$even dimension$-$4 operators of the Standard Model Extension (SME). We illustrate how bounds placed on these parameters can be associated with limits on the energy scale of the proposed neutrino interactions with cosmic scalars. Besides, as a case study, we employ a DUNE-like experimental configuration to assess the projected sensitivities to the $CPT-$even isotropic $c_{\alpha \beta}$ and $Z-$spatial $c_{\alpha \beta}^{ZZ}$ SME coefficients. For the case of the isotropic SME coefficients, an upper limit on the energy scale of the interaction can be placed. The current IceCube experiment and upcoming neutrino experiments such as DUNE, KM3NeT, IceCube-Gen2, and GRAND proposals, may clarify these classes of neutrino non-standard interactions.

\end{abstract}

\maketitle


\section{Introduction}
\label{intro}
The discovery of the Universe's accelerating expansion~\cite{SupernovaSearchTeam:1998fmf, SupernovaCosmologyProject:1998vns} is one of the most important, captivating, and puzzling open questions in cosmology~\cite{Sahni:2004ai,Alam:2004jy}. Several proposals to architecture Dark Matter (DM) and Dark Energy (DE) have been investigated in the literature. The Lambda Cold Dark Matter ($\Lambda$CDM) model is among the most popular explanations for cosmological observations, although recent results from the Dark Energy Spectroscopic Instrument (DESI)~\cite{DESI:2024mwx} show some tension with this model. Different candidates for DM particles have been considered, ranging in a very wide mass spectrum, making the Weakly Interacting Massive Particle (WIMP) paradigm one of the best motivated from a theoretical point of view~\cite{Arbey:2021gdg}. However, despite thorough searches for WIMPs, they all have failed to find any signature.

Other DM and DE candidates have also been studied in an effort to find a plausible explanation to the cosmological observations. For instance, ultralight scalars are well-motivated proposals for cosmological DM~\cite{Magana:2012ph, Ferreira:2020fam, Suarez:2013iw, Hui:2016ltb, Matos:2023usa}. Moreover, several ultralight scalars become apparent within the context of string theories~\cite{Arvanitaki:2009fg,Cicoli:2021gss, Acharya:2010zx,Marsh:2015xka}.~\footnote{Some strategies to hunt for ultralight scalars as cosmological DM involve atomic clocks~\cite{Arvanitaki:2014faa}, resonant-mass detectors~\cite{Arvanitaki:2015iga} and atomic gravitational wave detectors~\cite{Arvanitaki:2016fyj}.} Scalar fields minimally coupled to gravity are sufficiently justified models of DE~\cite{Li:2011sd}, namely quintessence~\cite{Wetterich:1987fm, Zlatev:1998tr}, symmetrons~\cite{Kading:2023hdb, Burrage:2018zuj}, and k-essence~\cite{Chiba:1999ka, Armendariz-Picon:2000nqq,Armendariz-Picon:2000ulo,Melchiorri:2002ux,Chiba:2002mw,Chimento:2003zf,Chimento:2003ta}.

In k-essence models, the scalar field plays a significant role in describing the DE puzzle. This field could have adequate behavior at early epochs and can reproduce the dynamical effects of the cosmological constant at late times. Classes of k-essence Lagrangians were introduced in several settings, for example, as a possible model for inflation~\cite{Armendariz-Picon:1999hyi, Garriga:1999vw}. Subsequently, k-essence models were used as another possibility to describe the characteristics of DE and as an alternative mechanism for unifying DE and DM~\cite{Scherrer:2004au}. Purely kinetic k-essence models~\cite{dePutter:2007ny, Gao:2010ia} are, in a way, as simple as quintessence models, because they rely only on one function ($F$) through the expression of the Lagrangian density ${\mathcal{L}} = F(X)$, where $X$ is the kinetic term.

Among other proposals, there are second-order derivative scalar field models, or generalized galileons~\cite{Nicolis:2008in, deRham:2010eu,Goon:2010xh,Goon:2011qf}, with the property that their equations of motion are second-order.

Interesting examples of this class of models are the so-called kinetic gravity braiding models, that are formulated from a Lagrangian that includes a D'Alambertian operator and an arbitrary function of a non-canonical kinetic term giving rise to appealing cosmological effects~\cite{Deffayet:2010qz,Pujolas:2011he, Kimura:2010di,Maity:2012dx}.

At the cosmological level, all the aforementioned models of DM and DE can be described as a perfect fluid through their energy-momentum tensor $T_{\varphi}^{\mu \nu}$~\cite{cosmoweinb}. At present, discrimination among them could be done at the level of the energy-momentum tensor perturbations.
On the other hand, the search for signatures confirming the existence of scalars is an important challenge. Proposals considering the interaction of scalar fields with neutrinos have demonstrated that a possible signature might be detected in long-baseline neutrino experiments~\cite{Berlin:2016woy, deSalas:2016svi, Krnjaic:2017zlz, Brdar:2017kbt, Smirnov:2019cae, Dev:2020kgz, Losada:2021bxx, Losada:2022uvr, Huang:2022wmz, Cordero:2022fwb, Sen:2023uga} or via modifications to the ultrahigh energy neutrino fluxes~\cite{Barranco:2010xt, Reynoso:2016hjr}. Besides, it has also been argued that interactions of neutrinos with scalars, if they exist, could induce an apparent violation of Lorentz and $CPT$ symmetries in the neutrino sector~\cite{Gu:2005eq, Ando:2009ts, Klop:2017dim, Capozzi:2018bps, Farzan:2018pnk, Ge:2019tdi, Gherghetta:2023myo, Arguelles:2023wvf, Arguelles:2023jkh, Lambiase:2023hpq, Cordero:2023hua, Arguelles:2024cjj}. These scalars can be identified as either DM or DE candidates. We refer to~\cite{Kostelecky:2003cr, Diaz:2016xpw, Torri:2020dec, Moura:2022dev, Barenboim:2022rqu} for comprehensive reviews concerning Lorentz and $CPT$ symmetry violations in the neutrino sector, within the Standard Model Extension (SME) framework~\cite{Colladay:1998fq}.
An initial proposal for $CPT$ violation in the neutrino sector was discussed in Ref.~\cite{Barenboim:2002tz}.

Recently, there has been an increased interest in searches for Lorentz invariance and $CPT-$breakdowns in neutrino oscillation experiments~\cite{Barenboim:2018ctx, IceCube:2021tdn, Sahoo:2021dit, Agarwalla:2023wft, Raikwal:2023lzk, Crivellin:2020oov, Testagrossa:2023ukh, Shukla:2024fnw}. For instance, comprehensive studies of the isotropic $CPT-$even SME coefficient, $(c_L)^{TT}$, at different long-baseline experiments can be found in Refs.~\cite{Agarwalla:2023wft, Raikwal:2023lzk}.~\footnote{As shown in~\cite{Diaz:2020aax}, it is possible to relate the isotropic SME coefficients $(c_L)^{TT}$ with the effects of the violation of the equivalence principle (VEP) in the neutrino sector. Besides, there is a correspondence between quantum-decoherence effects in neutrinos and the effective coefficients of the SME~\cite{Barenboim:2004wu, DeRomeri:2023dht, Barenboim:2024zfi}.} 
In Ref.~\cite{Delgadillo:2024bae}, we examine the phenomenology of isotropic and anisotropic $CPT-$odd SME coefficients, $(a_L)^T$ and $(a_L)^Z$, considering the Deep Underground Neutrino Experiment (DUNE)~\cite{DUNE:2020ypp} configuration.

In this paper, we consider the possibility of tensorial neutrino non-standard interactions with scalar fields; such as quintessence and other related DE models, as well as the case of ultralight dark matter (ULDM). We focus on the scenario where the $CPT-$even SME coefficients could correspond to a tensorial interaction of neutrinos with cosmological scalar fields. In this scenario, an apparent Lorentz$-$violating signature may manifest and give a sizable signal at upcoming and present neutrino oscillation experiments. Besides, as a case of study, we asses the projected sensitivities of a DUNE-like setup to these types of LIV scenarios. This is the first time sensitivities for the $CPT-$even SME coefficients $(c_L)^{ZZ}$ at DUNE are presented.

The manuscript is organized as follows: In Section~\ref{frame}, we introduce the theoretical foundations, where the basics of Lorentz invariance violations are presented together with the necessary Lagrangian formalism to describe the several models of scalar fields useful to model DM and DE. In Section~\ref{model}, we describe the phenomenology of a tensorial neutrino interaction with scalar fields as either DM or DE candidates and its connection with the $CPT-$even SME coefficients $(c_L)_{\alpha \beta}^{\mu \nu}$. In Section~\ref{numerics}, we assess the sensitivities to the $CPT-$even SME coefficients, $(c_L)_{\alpha \beta}^{\mu \nu}$, considering the DUNE configuration. Finally, we give our conclusions.

\section{Theoretical framework}
\label{frame}

Within the SME framework, violations of Lorentz invariance (LIV) in the fermion sector, are parameterized by the effective Lagrangian~\cite{Barenboim:2022rqu}
\begin{equation}
    \mathcal{L}_{ \text{eff}}^{\psi} = \mathcal{L}_{\text{SM}}^{\psi} + \mathcal{L}_{\text{LIV}} + \text{h.c.}\,,
\end{equation}
where $\mathcal{L}_{\text{SM}}^{\psi}$ is the Standard Model (SM) fermion Lagrangian, $\mathcal{L}_{\text{LIV}}$ being the Lorentz and $CPT-$violating Lagrangian term,
\begin{equation}
   - \mathcal{L}_{\text{LIV}}=  \frac{1}{2} \Big\{ p^{\mu}_{\alpha \beta}\Bar{\psi}_{\alpha} \gamma_{\mu} \psi_{\beta} + q^{\mu}_{\alpha \beta}\Bar{\psi}_{\alpha} \gamma_{5}\gamma_{\mu} \psi_{\beta} -i r^{\mu \nu}_{\alpha \beta}\Bar{\psi}_{\alpha} \gamma_{\mu}  \partial_{\nu}  \psi_{\beta}- i s^{\mu \nu}_{\alpha \beta}\Bar{\psi}_{\alpha} \gamma_{5}\gamma_{\mu}\partial_{\nu} \psi_{\beta} \Big\}\,.
\end{equation}
In the case of neutrinos, it is convenient to define
\begin{equation}
    (a_{L})^{\mu}_{\alpha \beta} = (p+q)^{\mu}_{\alpha \beta} \,\,\,\,\,{\rm and}\,\,\,\,\, (c_{L})^{\mu \nu}_{\alpha \beta} = (r+s)^{\mu \nu }_{\alpha \beta} \,,
\end{equation}
where the $(a_{L})^{\mu}_{\alpha \beta}$ coefficients are $CPT-$odd, while the $(c_{L})^{\mu \nu}_{\alpha \beta}$ coefficients are $CPT-$even. These type of $CPT-$conserving terms, $(c_{L})^{\mu \nu}_{\alpha \beta}$, were studied in the context of a Finslerian Geometrical model~\cite{Antonelli:2018fbv}, violations of diffeomorphism invariance~\cite{Reyes:2024hqi, Santos:2024iyc}, and neutrino interactions with a k-essence field~\cite{Gauthier:2009wc}. For other scenarios of neutrino$-$dark energy interactions, we refer the reader to Refs.~\cite{Khalifeh:2020bdg, Khalifeh:2021ree, Khalifeh:2021jfl}.

Lately, it has been suggested that violations of Lorentz invariance and $CPT$ symmetry in the neutrino sector may arise from a neutrino$-$current coupled with a dynamical cosmic field, $\varphi(t)$ (see, e.g., Refs.~\cite{Gu:2005eq, Ando:2009ts, Klop:2017dim, Capozzi:2018bps, Arguelles:2023jkh, Lambiase:2023hpq, Cordero:2023hua})
\begin{equation}
\label{livdark}
    (a_L)^{\mu}_{\alpha \beta} \Bar{\nu}_{\alpha} \gamma_{\mu}(1- \gamma_5) \nu_{\beta} \rightarrow y_{\alpha \beta} \frac{\partial^{\mu} \varphi}{\Lambda}  \Bar{\nu}_{\alpha} \gamma_{\mu}(1- \gamma_5) \nu_{\beta},
\end{equation}
being $y_{\alpha \beta}$ some coupling constants, $\Lambda$ the energy scale of the interaction, and $\varphi= \varphi(t)$ a time$-$varying scalar field, which could be identified as either a dark matter or dark energy candidate.~\footnote{For ultra-relativistic neutrinos, the phenomenology with axion-like dark matter is the same as the ultralight dark matter scenario~\cite{Lambiase:2023hpq, Gherghetta:2023myo, Huang:2018cwo}.} Here, violations of the Lorentz and $CPT$ symmetries emerge via the $CPT-$odd SME coefficients $(a_L)_{\alpha \beta}^{\mu} \rightarrow y_{\alpha \beta} \partial^{\mu} \varphi \Lambda^{-1}$.

Motivated by the aforementioned proposals, let us consider the corresponding free scalar field Lagrangian and its energy-momentum tensor, which are relevant for describing both, scalar field dark matter and dark energy quintessence~\cite{cosmoweinb, Ferreira:2020fam, Wetterich:1987fm},
\begin{equation}
  \begin{split}
    &~~~~ \mathcal{L}_{\varphi} = \frac{1}{2} g^{\mu \nu} \partial_{\mu} \varphi  \partial_{\nu} \varphi + V(\varphi),\\
    &~~~~ T^{\mu \nu}_{\varphi} =  \frac{2   }{ \sqrt{-g} } \frac{\delta}{\delta g_{\mu \nu}} \Big[\sqrt{-g}  \Big( \frac{1}{2} g^{\lambda \sigma} \partial_{\lambda} \varphi  \partial_{\sigma} \varphi + V(\varphi) \Big)  \Big]\,.
  \end{split}
\end{equation}
Therefore, the scalar field energy-momentum tensor $T^{\mu \nu}_{\varphi}$ can be expressed as
\begin{equation}\label{Eq:Tmnu1}
 T^{\mu \nu}_{\varphi}=  g^{\mu \nu}   \Big( \frac{1}{2} g^{\lambda \sigma} \partial_{\lambda} \varphi  \partial_{\sigma} \varphi + V(\varphi) \Big)  - g^{\lambda \mu } g^{\sigma \nu } \partial_{\lambda} \varphi  \partial_{\sigma} \varphi,
\end{equation}
and we identify the density and pressure of the scalar field $\varphi$ as
\begin{equation}
\begin{split}
   &\rho_{\varphi} = - \frac{1}{2} g^{\mu \nu} \partial_{\mu} \varphi  \partial_{\nu} \varphi + V(\varphi)\,\,\,\,\, {\rm and}\\
   &p_{\varphi}= - \frac{1}{2} g^{\mu \nu} \partial_{\mu} \varphi  \partial_{\nu} \varphi - V(\varphi)\,.
\end{split}    
\end{equation}
The scalar field four-velocity, $ U^{\mu}$, is given as
\begin{equation}
    U^{\mu}= \frac{ -\partial_{\mu}
    \varphi}{\sqrt{ -g^{\lambda \sigma} \partial_{\lambda} \varphi  \partial_{\sigma} \varphi }}\,.
\end{equation}
The isotropy and homogeneity of space-time require the stress-energy-momentum tensor for a free scalar field, $\varphi$, to be that of a perfect fluid, 
\begin{equation}
\label{perf}
\begin{split}
    &T^{ij}_{\varphi} = p_{\varphi} \delta^{i}_j; ~~T^{i0}_{\varphi} = T_{ \varphi0i} = 0;~~ T^{00}_{\varphi}  =  \rho_{\varphi}, \\
    &T^{\mu \nu}_{\varphi} = p_{\varphi} g^{ \mu \nu} + (p_{\varphi} +\rho_{\varphi}) U^{\mu} U^{\nu},
\end{split}
\end{equation}
subject to the constraint $g_{\mu \nu} U^{\mu} U^{\nu} = -1$. The conservation of $T^{\mu \nu}_{\varphi}$ in a Friedmann-Lemaitre-Robertson-Walker (FLRW) background leads to the continuity equation,
\begin{equation}
\label{continuity}
     \partial_0 T^{00}_{\varphi} = \dot{\rho}_{\varphi} + 3\frac{\dot{a}}{a} (p_{\varphi} +\rho_{\varphi}) = 0 \,,
\end{equation}
which gives the equation of motion of the scalar field,
\begin{equation}
\ddot{\varphi}+3H\dot{\varphi}+\frac{dV(\varphi)}{d\varphi}= 0 \,.
    \label{scalarevolution}
\end{equation}
Nevertheless, dark energy can be modeled by a k-essence field $\varphi$ which has very appealing properties and is described by the Lagrangian density $\mathcal{L}=K(\varphi,X)$ \cite{Armendariz-Picon:2000nqq} which is associated with the following energy-momentum tensor,
\begin{align}
\label{GENTENSOR1}
T_{\mu\nu}^{\varphi}=K,_{X} \nabla_{\mu} \varphi \nabla_{\nu} \varphi - g_{\mu\nu}K\,,
\end{align}
where $X=\nabla_{\alpha} \varphi \nabla^{\alpha} \varphi/2$ is the kinetic term and $K,_{X}=\partial K/ \partial X$. 
The energy density and pressure are given by
\begin{eqnarray}
\label{kess4}
\rho_{\varphi} = 2X K_{,X} - K \,\,\,\,\, {\rm and} \,\,\,\,\, &&  p_{\varphi} = K \, \, ,
\end{eqnarray}
respectively, and the equation of motion can be obtained from the continuity equation,
\begin{equation}
  (2XK_{,XX}+K_{,X}) \ddot{\varphi} + K_{,X\varphi}\dot{\varphi}^2 +3H K_{,X}\dot{\varphi}= K_{,\varphi} \, ,
  \label{eqmotionk}
\end{equation}
where $K_{,X\varphi} = \partial^2K/\partial X \partial \varphi$ and $K_{,\varphi}=\partial K/ \partial \varphi$.

Another possible interesting scalar field used to describe the dark energy evolution is the braided scalar field described by the Lagrangian:
\begin{align}
\label{BR1}
\mathcal{L}=\square\; \varphi G\left(X,\varphi\right),
\end{align}
where $\square \;
\varphi=\nabla^{\mu}\nabla_{\mu}\varphi$ and the function $G$ is arbitrary \cite{Pujolas:2011he}. 
The energy-momentum tensor, $T_{\mu\nu}^{\varphi}$, in covariant form, is given by
\begin{align}
    \label{BR5}
    T_{\mu\nu}^{\varphi} = \frac{2}{\sqrt{-g}}\frac{\delta S_{\varphi}}{\delta g^{\mu\nu}} = \mathcal{L}_{X}\nabla_{\mu} \varphi \nabla_{\nu} \varphi - g_{\mu\nu} P_{\varphi} - \nabla_{\mu} G \nabla_{\nu} \varphi - \nabla_{\nu} G \nabla_{\mu} \varphi\,\,\, ,
\end{align}
where $P_\varphi = \nabla^{\lambda} \varphi \nabla_{\lambda} G_\varphi$, and $G_\varphi = \partial G/\partial \varphi$. This energy-momentum tensor can be described in terms of an imperfect fluid \cite{Pujolas:2011he} . We start defining some quantities to describe relativistic fluids. A local rest frame is set defining the normalized four-velocity $U_{\mu}$
\begin{align}
    \label{BR21}
    U_{\mu} \equiv \frac{\nabla_{\mu}\varphi}{\sqrt{2X}}, \quad U_{\mu}U^{\mu}=1, \\ 
    \label{BR22}
    a_{\mu} \equiv  U^{\nu} \nabla_{\nu}U_{\mu},
\end{align}
with $a_{\mu}$ is the four-acceleration which is orthogonal to velocity, $U_{\mu} a^{\mu} = 0$.
The expansion, $\vartheta$, and the diffusivity, $\Omega$, are written as
\begin{eqnarray}
    \label{BR24}
    \vartheta = \nabla_{\mu} U^{\mu}, \quad  \Omega = 2 X G_{X},
\end{eqnarray}
where $G_{X}=\partial G/ \partial X$. The energy-momentum tensor also can be expressed in this way: 
\begin{align}
    \label{BR11}
    T_{\mu\nu}^{\varphi} = \rho_{\varphi} U_{\mu} U_{\nu} - \perp_{\mu\nu} p_{\varphi} + U_{\mu} q_{\nu} + U_{\nu} q_{\mu},
\end{align}
where $m = \sqrt{2X} = \dot{\varphi}$ is the chemical potential, $q_{\mu} = -m \Omega a_{\mu}$ is the heat flux (purely spatial, $U_{\mu} q^{\mu} = 0$) and $\perp_{\mu\nu} = g_{\mu\nu} - U_{\mu}U_{\nu}$ is the transverse projector. The energy density and isotropic pressure are: 
\begin{eqnarray}
    \label{BR9}
    \rho_{\varphi} & \equiv & T_{\mu\nu}^{\varphi} U^{\mu} U^{\nu} = -2 X G_{\varphi} + \vartheta m \Omega, \\ \nonumber \\
    \label{BR10}
    p_{\varphi} & \equiv & -\frac{1}{3}T^{\mu\nu}_{\varphi} \perp_{\mu\nu} = -2 X G_{\varphi} - \Omega \dot{m}.
\end{eqnarray}
The conservation of $T_{\mu\nu}^{\varphi}$ using these definitions can be expressed as:  
\begin{align}
    \label{BR12}
    U_{\nu} \nabla_{\mu} T^{\mu\nu}_{\varphi} = \dot{\rho_{\varphi}} + \vartheta \left(\rho_{\varphi}  + p_{\varphi} \right) - \nabla_{\lambda} \left( m \Omega a^{\lambda} \right) + m \Omega a_{\lambda}a^{\lambda} = 0.
\end{align}
The equation of motion for $\varphi$ can be obtained from the last equation
\begin{align}
    \label{BR14}
    \nabla_{\mu}\left[ 2 G_{\varphi} \nabla^{\mu} \varphi  - \square \varphi G_{X} \nabla^{\mu}\varphi + G_{X}\nabla^{\mu}X \right]  
    = \nabla^{\lambda} \varphi \nabla_{\lambda} G_{\varphi} .
    \end{align}
In the cosmological background, the evolution of the scalar field reduces to the dynamics of a perfect fluid which is described only through its energy density and pressure.

\section{Tensorial neutrino LIV}
\label{model}

Consider the effective Lagrangian (see Appendix \ref{appx} for further details)
\begin{equation}
   \label{efftensor}
    -\mathcal{L}_{\text{eff}} = -i \frac{\lambda_{\alpha \beta} }{M_*^4 }T^{\mu \nu}_{\varphi} \bar{\nu}_{\alpha}\gamma_{\mu}(1-\gamma_5) \partial_\nu \nu_{\beta},
\end{equation}
being $\lambda_{\alpha \beta}$ a coupling constant matrix, $M_*$ the energy scale of the interaction, and $T^{\mu \nu}_{\varphi}$ is the tensor associated to the scalar field $\varphi$, for instance those of Eqs.~(\ref{Eq:Tmnu1}), (\ref{GENTENSOR1}), and (\ref{BR5}).
Hence, we can identify the $CPT-$even LIV coefficients of the SME $(c_{L})^{\mu \nu}_{\alpha \beta} = c^{\mu \nu}_{\alpha \beta}$ as
\begin{equation}
 c^{\mu \nu}_{\alpha \beta} \rightarrow \frac{\lambda_{\alpha \beta} }{M_*^4 }T^{\mu \nu}_{\varphi}.  
\end{equation}
Here, the scalar field $\varphi$ could be one of the DM or DE candidates described in Section~\ref{frame}.
Besides, the effective interaction (Eq.~\ref{efftensor}) may induce potential scattering among the neutrinos and the scalar particles. However, such interactions would be expected to be negligible (see Appendix B of Ref.~\cite{Klop:2017dim}).
Henceforth, we discuss the case of a tensorial neutrino-scalar field interaction.~\footnote{A study of the the back-reaction effects from the tensorial neutrino-scalar field interaction is beyond the scope of this work.}

\subsection{Isotropic LIV coefficients $c_{\alpha \beta}$ from a neutrino-scalar field interaction}
\label{isotropic}
From the effective Lagrangian in Eq.~(\ref{efftensor}), the corresponding isotropic $CPT$-even LIV coefficients $c_{\alpha \beta}$ are
\begin{equation}
\label{isoc}
c_{\alpha \beta}= c_{\alpha \beta}^{00} \rightarrow \frac{  \lambda_{\alpha \beta} }{M_*^4  }T^{0 0}_{\varphi} =\frac{ \lambda_{\alpha \beta} }{M_*^4  } \rho_{\varphi},  
\end{equation}
considering the scalar field $\varphi$ as ultralight dark matter (ULDM)~\cite{Suarez:2013iw, Ferreira:2020fam}, with corresponding local DM density in the Milky Way, $\rho_{\varphi, \odot} \sim \rho_{\text{DM}, \odot} \simeq 2 \times 10^{-6}$ eV$^4$~\cite{deSalas:2019pee, deSalas:2020hbh, Sivertsson:2022riu}, an accelerator-based experiment similar to DUNE, with sensitivity $c_{\alpha \beta} \sim [1-10]\times 10^{-25}$ (left panel of Fig.~\ref{f1dune} in Sec.~\ref{numerics}), could potentially probe an energy scale of the interaction
 \begin{equation}
   M_* \sim [3 - 6] \times 10^{4}~\text{eV}~\big( \lambda_{\alpha \beta}/ \mathcal{O}(1)  \big)~\big(\rho_{\varphi, \odot} / 10^{-6}~\text{eV}^4\big).
\end{equation}
On the other hand, if the scalar field $\varphi$ is considered to be a DE candidate, namely quintessence, the corresponding dark energy density $\rho_{\varphi}^{\text{DE}} \simeq 3 \rho_{\text{DM}}^{\text{avg}} \sim 10^{-10}$ eV$^4$, in this scenario, the DUNE setup could potentially probe an energy scale
\begin{equation}
   M_* \sim [5 - 10] \times 10^{3}~\text{eV}~\big( \lambda_{\alpha \beta}/ \mathcal{O}(1)  \big)~\big(\rho_{\varphi}^{\text{DE}} / 10^{-10}~\text{eV}^4\big).
\end{equation}
At the cosmological level, the aforementioned models of dark energy outlined in Section~\ref{frame}, can be described by a dynamical scalar field $\varphi(t)$, thus, for all cases, they would predict $T^{00}_\varphi = \rho_{\varphi}^{\text{DE}}$.
However, limits from astrophysical neutrinos ($60~\text{TeV}\lesssim E_\nu \lesssim$ PeV) employing the IceCube astrophysical neutrino flavour data-set constraint $c_{\alpha \beta} \lesssim 10^{-34}$~\cite{IceCube:2021tdn} (such limits are expected to be improved by the combination of a two-detector fit, namely IceCube-Gen2 and GRAND~\cite{Testagrossa:2023ukh}), therefore~\footnote{Besides, in the standard model, neutrinos are part of the $SU(2)_L$ doublet, hence, we could have potentially induced LIV effects for the charged leptons. However, for the case of electrons, limits from astrophysical observations constraint $c_{ee}\in(-80~\text{to}~4)\times10^{-20}$~\cite{Kostelecky:2008ts}, which are several orders of magnitude weaker than the constraints derived in the neutrino sector $c_{\alpha \beta} \lesssim 10^{-34}$~\cite{IceCube:2021tdn}.} 
\begin{equation}
 c_{\alpha \beta} \sim \frac{ \lambda_{\alpha \beta} }{M_*^{4}  } \rho_{\varphi}  \lesssim 10^{-34}~,
\end{equation}
considering $\varphi$ as DM, ultra-high energy (UHE) neutrino experiments could potentially be sensitive to an energy scale
\begin{equation}
   M_* \gtrsim 10^7~\text{eV}~\big( \lambda_{\alpha \beta}/ \mathcal{O}(1)  \big)~\big( \rho_{\varphi, \odot} / 10^{-6}~\text{eV}^4\big)~\big(c_{\alpha \beta} / \lesssim 10^{-34}\big)~,
\end{equation}
on the other hand, considering $\varphi$ as DE, UHE neutrino experiments could potentially be sensitive to an energy scale
\begin{equation}
  M_* \gtrsim 10^6~\text{eV}~\big( \lambda_{\alpha \beta}/ \mathcal{O}(1)  \big)~\big( \rho_\varphi^{\text{DE}} / 10^{-10}~\text{eV}^4\big)~\big(c_{\alpha \beta} / \lesssim 10^{-34}\big).
\end{equation}
In Table.~\ref{tab:2}, we display the energy scale reach $M_*$ at IceCube as well as other neutrino experiments. Besides, considering only the derivative coupling on the scalar fields (see Appendix~\ref{appx}), unitarity bounds in collisions with quarks at the LHC set $M_*\gtrsim 30$ GeV, while unitarity bounds from LEP when colliding electrons and positrons impose $M_* \gtrsim 3$ GeV~\cite{Brax:2014vva}.

\subsection{Directional dependent LIV coefficients from a neutrino-scalar field interaction}
\label{directional}
Searches of directional dependent LIV effects can be performed at neutrino experiments such as the KM3NeT neutrino telescope and the IceCube neutrino observatory, with neutrino energies $E_\nu \gtrsim 10^5$ GeV~\cite{Klop:2017dim, Telalovic:2023tcb}.
For instance, regarding directional dependent Lorentz violating effects in the $Z-$direction, the projected sensitivities of a neutrino long-baseline experiment similar to DUNE to the $CPT-$even $Z-$spatial LIV coefficients are, $c_{\alpha \beta}^{ZZ} \sim [1-10]\times 10^{-24}$ (right panel of Fig.~\ref{f1dune} in Sec.~\ref{numerics}) and $c_{\alpha \alpha}^{ZZ}-c_{\tau \tau}^{ZZ} \simeq [1-10] \times 10^{-24}$ (right panel of Fig.~\ref{f2dune} in Sec.~\ref{numerics}), both at 95\% C.L., accordingly.

From the phenomenological Lagrangian in Eq.~(\ref{efftensor}), the corresponding $CPT-$even $Z-$spatial LIV coefficients in terms of the scalar field tensor are
\begin{equation}
\label{spatialc}
 c_{\alpha \beta}^{ZZ} \rightarrow \frac{ \lambda_{\alpha \beta} }{M_*^{4}  } (T^{ZZ}_{\varphi}+ \delta T^{ZZ}_{\varphi}),  
\end{equation}
where $\delta T_{\varphi}^{\mu \nu}$ accounts for the perturbations of the scalar field tensor $T^{\mu \nu}_{\varphi}$. However, perturbations for the several dark energy models outlined: quintessence, k-essence, or kinetic gravity braiding models, are expected to be different.~\footnote{A detail study of the energy-momentum perturbations for these models is beyond the scope of this paper and we leave it for a future work.} Subsequently, we refer to the dark energy quintessence model for simplicity. Still, a similar phenomenology applies to the other dark energy models.

The perturbed stress-energy-momentum tensor for the scalar field $\varphi(t)$, with perturbation $\delta \varphi (t, \textbf{x})$ in the FLRW background with a Newtonian gauge is given as~\cite{Magana:2012ph}
\begin{equation}
     \delta T^{ij}_{\varphi} = \delta p_{\varphi} \delta^{i}_j; ~~\delta T^{i0}_{\varphi} = - \frac{1}{a} \big( \dot{\varphi}(t) \partial_{i}( \delta \varphi) \big);~~\delta T^{00}_{\varphi}  =  \delta \rho_{\varphi}.
\end{equation}
Considering $\varphi$ as ULDM, from the equation of state of the scalar field $p_\varphi = \omega \rho_\varphi$, with $\omega =0$, the leading contribution $T^{ZZ}_{\varphi}=0$, hence, in this case the $CPT-$even $Z-$spatial LIV coefficients are
\begin{equation}
 |c_{\alpha \beta}^{ZZ}| \rightarrow \frac{  \lambda_{\alpha \beta} }{M_*^{4}  }  |\delta T^{ZZ}_{\varphi} | \sim \frac{  \lambda_{\alpha \beta} }{M_*^{4} } \big| \big( \dot{\varphi}(t) \dot{\delta \varphi}(t,\textbf{x}) -\dot{\varphi}^2(t) \Phi(t,\textbf{x}) -V(\varphi)_{,\varphi}~\delta\varphi(t,\textbf{x}) \big)\big|, \end{equation}
here $\Phi(t,\textbf{x})$ plays the role of the gravitational potential, while $\delta \varphi (t,\textbf{x})$ is the scalar field perturbation~\cite{Magana:2012ph}. On the other hand, if $\varphi$ is a DE candidate (quintessence) from the equation of state of the scalar field $p_\varphi = \omega \rho_\varphi^{\text{DE}}$, with $\omega =-1$, 
\begin{equation}
 |c_{\alpha \beta}^{ZZ}| \rightarrow \frac{\lambda_{\alpha \beta} }{M_*^{4}  }  |T^{ZZ}_{\varphi} + \delta T^{ZZ}_{\varphi} | \sim \frac{  \lambda_{\alpha \beta} }{M_*^{4}  }  \rho_\varphi^{\text{DE}}.  
\end{equation}
For instance, in Table.~\ref{tab:2}, we show the energy scale reach $M_*$ within this scenario, considering limits set on the $CPT-$even SME coefficients ($c_{\alpha \beta}^{ZZ}$) from several neutrino experimental configurations. In a similar fashion, considering $\varphi$ as either ULDM or DE quintessence, the leading contribution $T^{TZ}_{\varphi}=0$ (Eq.~\ref{perf}), therefore, the $CPT-$even coefficients from the $TZ-$sector can be described as  
\begin{equation}
 |c_{\alpha \beta}^{TZ}| \rightarrow \frac{\lambda_{\alpha \beta} }{M_*^4 }  |\delta T^{TZ}_{\varphi} | \sim \frac{ \lambda_{\alpha \beta} }{M_*^{4} } \Big|-\frac{1}{a} \big( \dot{\varphi}(t) \frac{\partial  }{\partial z} \delta \varphi(t, x, y, z) \big) \Big|. 
\end{equation}

\section{Numerical analysis and expected sensitivities}
\label{numerics}
Long-baseline neutrino oscillation experiments play a significant role in both deciphering mysteries within the traditional three-neutrino oscillation picture and exploring other novel physics scenarios, including the potential breaking of the Lorentz and $CPT$ symmetries. Hence, as a case of study, we will focus on a long-baseline experimental configuration, the Deep Underground Neutrino Experiment (DUNE), which is a next-generation accelerator-based neutrino oscillation experiment that will consist of up to 40 kt of liquid argon (far) detector located at the Sanford underground research facility (SURF) in South Dakota~\cite{DUNE:2020jqi}. Moreover, this configuration is expected to deliver a neutrino flux with a mean neutrino energy $E_\nu \sim$ 3 GeV, which will be located at a distance of $L \sim$ 1300 km from the beam source (on-axis) at Fermilab (we refer the reader to Refs.~\cite{Agarwalla:2023wft, DUNE:2020ypp}, for a detailed discussion regarding the experimental configuration).

In order to obtain sensitivities to the LIV coefficients at DUNE, we use the \textsc{GLoBES} software~\cite{Huber:2004ka, Huber:2007ji} and its additional NSI tool \emph{snu.c}~\cite{Kopp:2006wp, Kopp:2007rz} which was modified to implement the $CPT-$even coefficients of the SME at the Hamiltonian level. Moreover, to simulate the DUNE configuration, we employ the available~\textsc{GLoBES} ancillary files~\cite{DUNE:2021cuw} and specifications from the Technical Design Report (TDR) configuration~\cite{DUNE:2020ypp}. Furthermore, in this work, we have contemplated a $10-$year running time, evenly distributed among neutrino and anti-neutrino modes. To simulate the DUNE event spectra, we consider the reconstructed neutrino and anti-neutrino energy range from 0 to 18 GeV for both appearance and disappearance channels. While elaborating our sensitivity plots, we performed a full spectral analysis with a total of 70 bins in the aforementioned energy range (having non-uniform bin widths). We have 64 bins each having a width of 0.125 GeV in the energy range of 0 to 8 GeV, and 6 bins with variable widths beyond 8 GeV~\cite{DUNE:2021cuw}.  

In this analysis, we have considered the following Hamiltonian
\begin{equation}
    H=H_0 + H_{\text{MSW}} + H_{\text{LIV}},
\end{equation}
here, $H_0$ and $H_{\text{MSW}}$ are the standard neutrino Hamiltonian in vacuum and matter, respectively. Besides, $H_{\text{LIV}}$ is the contribution from the Lorentz invariance violation (LIV) sector, which can be parameterized as~\cite{Kostelecky:2004hg, Diaz:2009qk, Mishra:2023nqf}
\begin{equation}
\label{LIVHAM}
 H_{\text{LIV}} =  -\frac{E_\nu}{2} \big[ (3-\hat{N}_Z^2)(c_L)_{\alpha \beta}^{TT} +(3\hat{N}_Z^2-1)(c_L)_{\alpha \beta}^{ZZ} -2\hat{N}_Z (c_L)_{\alpha \beta}^{TZ} \big],
\end{equation}
where $E_\nu$ is the neutrino energy, $(c_L)_{\alpha \beta}^{\mu \nu}$ are the $CPT-$even LIV coefficients of the SME (being $\alpha, \beta = e, \mu, \tau$, and $\mu, \nu = T, X, Y, Z$), for non-diagonal $\alpha \neq \beta$, $c_{\alpha \beta}^{\mu \nu} = |c_{\alpha \beta}^{\mu \nu}| e^{i\phi_{\alpha \beta}^{\mu \nu}}$ (from now on, we will denote $(c_L)^{\mu \nu}_{\alpha \beta} = c_{\alpha \beta}^{\mu \nu}$, and $(c_L)^{TT}_{\alpha \beta} = c_{\alpha \beta}$~), and  
\begin{equation}
    \hat{N}^Z = -\sin\chi\sin\theta\cos\phi+\cos\chi\cos\theta,
\end{equation}
is the spatial $Z-$component factor, expressed in terms of local spherical coordinates at the detector, that represents the direction of neutrino propagation in the Sun-centered frame. Being $\chi$ the colatitude of the detector, $\theta$ the angle at the detector between the beam direction and vertical, and $\phi$ the angle between the beam and east of south~\cite{Kostelecky:2004hg}. In the particular case of the DUNE location, $\hat{N}^Z \simeq 0.16$~\cite{Delgadillo:2024bae}.

\subsection{Sensitivity to $CPT-$even SME coefficients}
\label{method}
In order to assess the statistical significance to $CPT-$even SME coefficients, we employ a chi-squared test, we have considered the muon neutrino disappearance channel $P(\nu_\mu \rightarrow \nu_\mu)$ as well as the electron neutrino appearance channel $P(\nu_\mu \rightarrow \nu_e)$, from a muon-neutrino beam with neutrino and antineutrino data sets.~\footnote{A detailed discussion of the impact of the $CPT-$even SME coefficients in the neutrino oscillation probability is beyond of the scope of this work.} The total $\chi^2-$function is provided as in Ref.~\cite{Cordero:2022fwb} 
\begin{equation}
    \chi^2 = \sum_{k} \tilde{\chi}^2_{k} +\chi^2_{\text{prior}},
\end{equation}
where the corresponding $\tilde{\chi}^2_{k}-$function for each channel ($k$), appearance or disappearance is given as in Ref.~\cite{Huber:2002mx}
\begin{equation}
\begin{split}
    &\tilde{\chi}_{k}^2= \min_{\zeta_{j}} \Bigg[  \sum _{i}^{n_{\text{bin}}} 2 \Bigg\{ N_{i,\text{test}}^{3 \nu+\text{LIV}}( \Pi, \Gamma, \{\zeta_{j}\})-N_{i,\text{true}}^{3\nu} +  N_{i,\text{true}}^{3\nu} \log \frac{N_{i,\text{true}}^{3\nu}}{N_{i,\text{test}}^{3 \nu+\text{LIV}}( \Pi, \Gamma, \{\zeta_{j}\})} \Bigg\} \\
    &~~~~~~~~~~~~~~~ + \sum_{j}^{n_{\text{syst}}} \Big(\frac{\zeta_{j}}{\sigma_{j}}\Big)^2 \Bigg],
\end{split}
\end{equation}
here, $N_{i,\text{true}}^{3\nu}$ refers to the simulated events at the $i$-th energy bin (considering the standard three neutrino oscillations picture), while $N_{i,\text{test}}^{3\nu +\text{LIV}}( \Pi, \Gamma, \{\zeta_{j}\})$ are the computed events at the $i$-th energy bin including $CPT-$even SME coefficients (one parameter at a time). 
In addition, $\Pi = \{\theta_{12}, \theta_{13}, \theta_{23}, \delta_{CP}, \Delta m_{21}^2, \Delta m^2_{31}\}$ is the set of neutrino oscillation parameters, while $\Gamma = \{|c_{\alpha \beta}|, \phi_{\alpha \beta}, c_{\alpha \alpha},~\cdots, |c_{\alpha \beta}^{ZZ}|, \phi_{\alpha \beta}^{ZZ}, c_{\alpha \alpha}^{ZZ}, \cdots  \}$ is the set of either isotropic ($c_{\alpha \beta}$), $Z-$spatial ($c_{\alpha \beta}^{ZZ}$) or ($c_{\alpha \beta}^{TZ}$) SME coefficients, where $\{\zeta_{j}\}$ are the nuisance parameters to account for the systematic uncertainties. Furthermore, $\sigma_{j}$ are the systematic uncertainties as reported in the DUNE TDR~\cite{DUNE:2020ypp}. Besides, to obtain our simulated events; we have considered the corresponding neutrino oscillation parameters as~\emph{true} values, namely $\Delta m^{2}_{21}=7.5 \times 10^{-5}~\text{eV}^{2}$, $\Delta m^{2}_{31} = 2.55 \times 10^{-3}~\text{eV}^2$, $\theta_{12} = 34.3^{\circ}$, $\theta_{13} = 8.53^{\circ}$, $\theta_{23} = 49.26^{\circ}$, and $\delta_{CP} =1.08 \pi$, corresponding to the best-fit values with normal mass ordering (NO) from Ref.~\cite{deSalas:2020pgw} as displayed in Tab.~\ref{tab:1}.

\begin{table}[ht]
\caption{\label{tab:1}Standard oscillation parameters used in our analysis~\cite{deSalas:2020pgw}. We consider the normal mass ordering (NO) throughout this study.}
\centering
\begin{tabular}{c  c}
\hline \hline
Oscillation parameter \,\,& best-fit \textbf{NO}  \\
\hline 
$\theta_{12}$ & 34.3$^{\circ}$ \\
$\theta_{23}$ & 49.26$^{\circ}$ \\
$\theta_{13}$ &  8.53$^{\circ}$ \\
$\Delta m^2_{21}$ [10$^{-5}$~eV$^2$] & 7.5  \\ 
$|\Delta m_{31}^2|$ [10$^{-3}$~eV$^2$] & 2.55  \\ 
$\delta_{CP}/ \pi$ & 1.08  \\
\hline \hline
\end{tabular}
\end{table} 
Furthermore, to include external input and marginalization for the standard oscillation parameters (in the total $\chi^2-$function), Gaussian priors~\cite{Huber:2002mx} are utilized,
\begin{equation}
    \chi^2_{\text{prior}}= \sum_{p}^{n_{\text{priors}}}   \frac{\big(\Pi_{p,\text{true}}-\Pi_{p,\text{test}}\big)^2}{\sigma^2_{p}}, 
\end{equation}
the central values of the oscillation parameter priors 
($\Pi_{p, \text{true}}$) were fixed to their best-fit~\cite{deSalas:2020pgw}, considering the normal mass ordering, $\sigma_p$ is the uncertainty on the oscillation prior, which corresponds to a 1$\sigma$ error (68.27\% confidence level C.L.\,).
\begin{figure}[H]
\begin{subfigure}[h]{0.49\textwidth}
			\caption{  }
			\label{f1a}
\includegraphics[width=\textwidth]{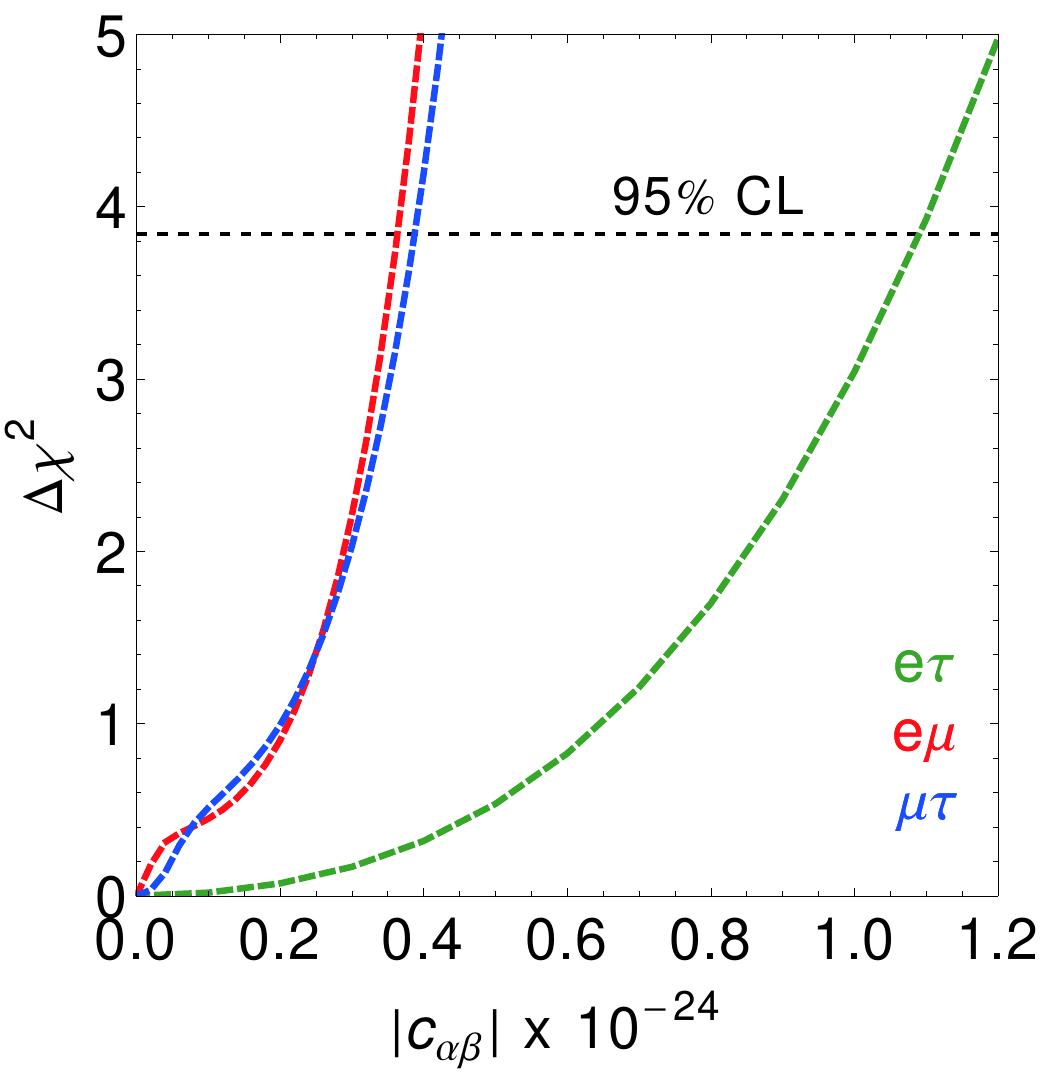}
		\end{subfigure}
		\hfill
		\begin{subfigure}[h]{0.48\textwidth}
			\caption{}
			\label{f1b}
			\includegraphics[width=\textwidth]{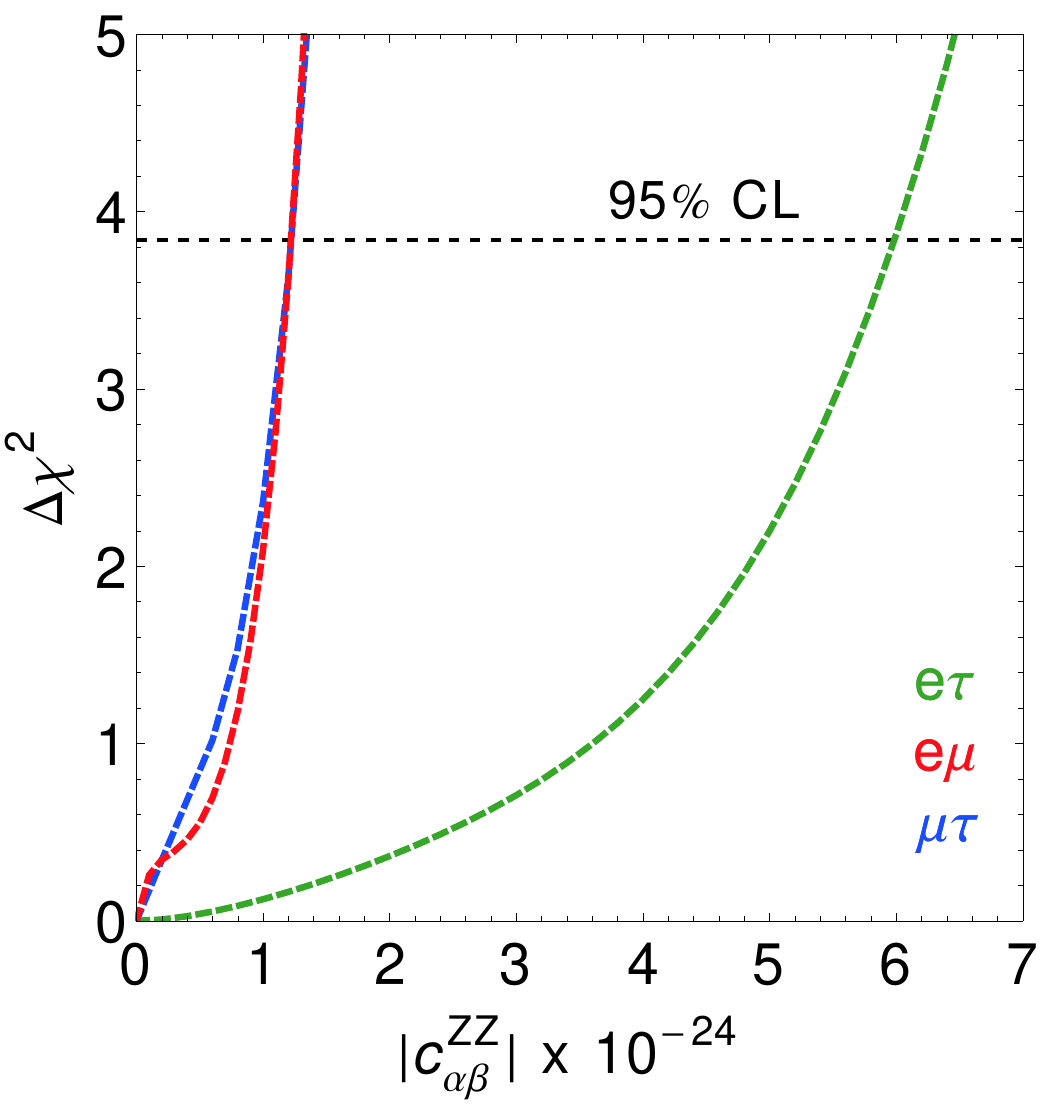}
		\end{subfigure}
		\hfill	
		 \caption{Projected 95\% C.L. sensitivities to the SME coefficients $|c^{\mu \nu}_{\alpha\beta}|$ at the DUNE (TDR setup). Here, we have marginalized over the corresponding LIV phases $\phi_{\alpha \beta}$ and $\phi_{\alpha \beta}^{ZZ}$ from [0$-2\pi$], as well as $\theta_{23}$ and $\delta_{CP}$, considering a 1$\sigma$ uncertainty of 10\% and 15\%, respectively. All the remaining oscillation parameters were fixed to their NO best fit values~\cite{deSalas:2020pgw}. Refer to the text for details.}
  \label{f1dune}
\end{figure}

In Fig.~\ref{f1dune}, we display our results of the expected 95\% C.L. sensitivities to the non-diagonal SME coefficients, namely the isotropic $|c_{\alpha \beta}| \lesssim 1.1 \times 10^{-24}$ (left panel) and $Z-$spatial $|c_{\alpha \beta}^{ZZ}| \lesssim 6.2 \times 10^{-24}$ (right panel), for the DUNE configuration. Here, we have marginalized over the corresponding LIV phases $\phi_{\alpha \beta}$ and $\phi_{\alpha \beta}^{ZZ}$ from [0$-2\pi$], as well as the atmospheric mixing angle $\theta_{23}$ and the leptonic phase $\delta_{CP}$, considering a 1$\sigma$ uncertainty of 10\% and 15\%~around their NO best-fit values, as displayed in Table~\ref{tab:1}. All the remaining oscillation parameters were fixed to their best fit value with normal mass ordering~\cite{deSalas:2020pgw}. Our results of the projected sensitivities to the isotropic LIV coefficients $\abs{c_{\alpha \beta}}$, are in agreement with those from Refs.~\cite{Agarwalla:2023wft, Raikwal:2023lzk}. However, main differences arising on the total number of bins and energy range selected for the reconstructed neutrino energies, as well as the neutrino oscillation benchmark values considered in the calculation of the expected number of events.

To put our results in perspective, existing bounds from the Super-Kamiokande experiment constraint $|c_{e \mu}| < 8.0 \times 10^{-27}$ and $|c_{e \tau}| < 9.3 \times 10^{-25}$ at 95$\%$ C.L.~\cite{Super-Kamiokande:2014exs}, while IceCube sets $\text{Re}(c_{\mu \tau}) < 7 \times 10^{-34}$ with a Bayes factor $>31.6$~\cite{IceCube:2021tdn}. Furthermore, a test for Lorentz and $CPT$ violation with the MiniBooNE low-energy excess restrict $c_{e \mu}^{ZZ}< (2.6 \pm 0.8)\times 10^{-19}$~\cite{Katori:2010nf, Kostelecky:2008ts}, while limits from the Double Chooz experiment set Re$(c^{ZZ}_{e \tau})< 4.9 \times 10^{-17}$~\cite{Katori:2013jca}. For projected sensitivities and some experimental constraints on the LIV coefficients, $c_{\alpha \beta}$ and $c_{\alpha \beta}^{ZZ}$, see Tab.~\ref{tab:2}.

Regarding the diagonal SME coefficients $c_{\alpha \alpha}-c_{\tau \tau}$ and $c_{\alpha \alpha}^{ZZ}-c_{\tau \tau}^{ZZ}$, we have fixed $c_{\tau \tau}=0$, as well as $c_{\tau \tau}^{ZZ} = 0$, since we can redefine the diagonal elements up to a global constant. 

\begin{figure}[H]
\begin{subfigure}[h]{0.49\textwidth}
			\caption{  }
			\label{f2a}
\includegraphics[width=\textwidth]{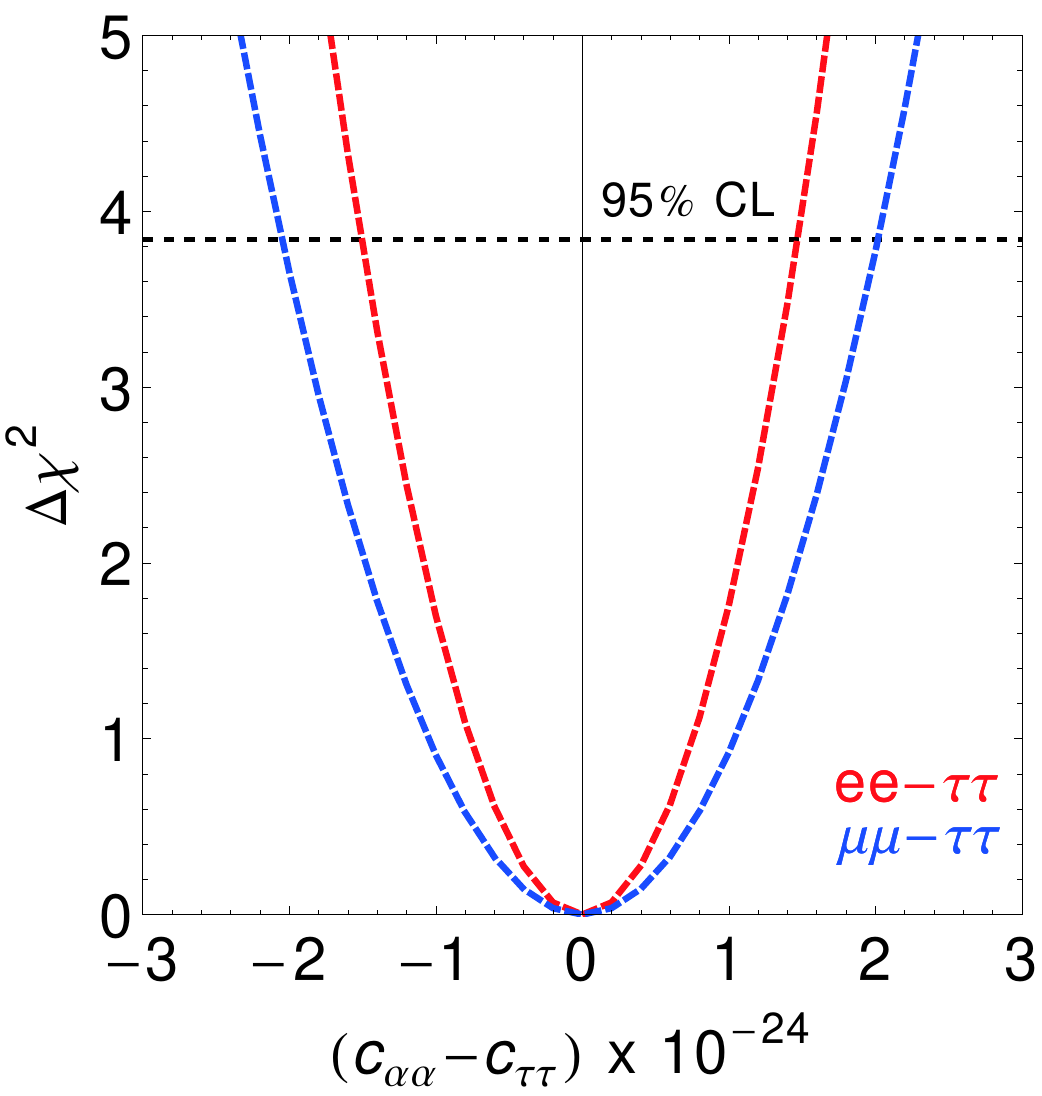}
		\end{subfigure}
		\hfill
		\begin{subfigure}[h]{0.49\textwidth}
			\caption{}
			\label{f2b}
			\includegraphics[width=\textwidth]{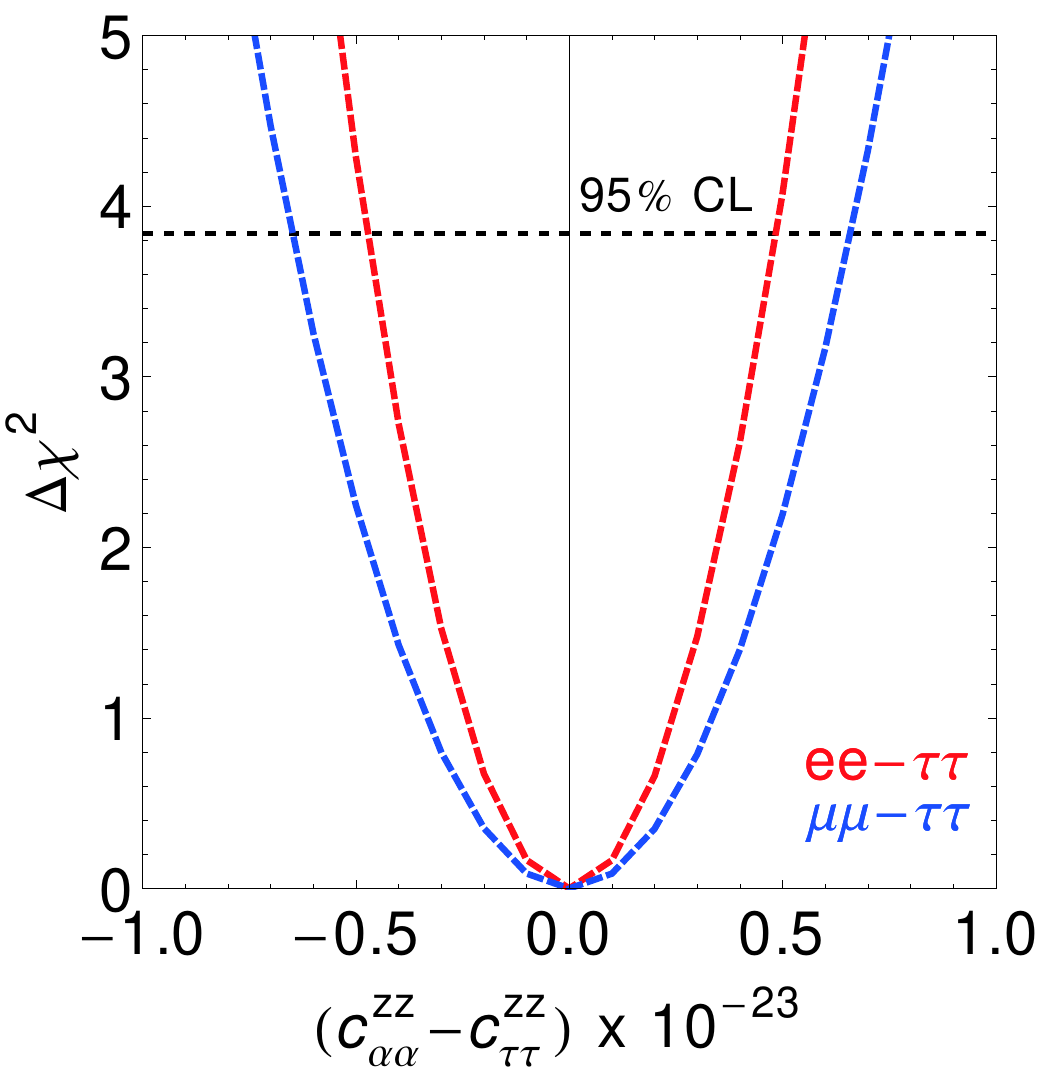}
		\end{subfigure}
		\hfill	
		 \caption{Expected 95\% C.L. sensitivities to the diagonal SME coefficients $c^{\mu \nu}_{\alpha \alpha}-c^{\mu \nu}_{\tau \tau}$ at the DUNE (TDR setup). Here, we have marginalized over the corresponding mixing angle $\theta_{23}$ and $\delta_{CP}$, considering a 1$\sigma$ uncertainty of 10\% and 15\%, respectively. All the remaining oscillation parameters were fixed to their NO best fit values~\cite{deSalas:2020pgw}. Refer to the text for details.}
  \label{f2dune}
\end{figure}
Our results of the projected 95\% C.L. sensitivities to the $c_{\alpha \alpha}-c_{\tau \tau}$ (left panel) and $c_{\alpha \alpha}^{ZZ}-c_{\tau \tau}^{ZZ}$ (right panel) coefficients at the DUNE experiment are shown in Fig.~\ref{f2dune}. We have marginalized over the corresponding atmospheric mixing angle $\theta_{23}$ and Dirac $CP$-violating phase $\delta_{CP}$, considering a 1$\sigma$ uncertainty of 10\% and 15\%~around their NO best-fit values, as show in Table~\ref{tab:1}. All the remaining oscillation parameters were fixed to their best fit value with normal mass ordering~\cite{deSalas:2020pgw}. We observe that the DUNE will be able to set limits on the isotropic $CPT$-even SME coefficients $c_{e e}-c_{\tau \tau} < 1.4 \times 10^{-24}$ and $c_{\mu \mu}-c_{\tau \tau} < 2.3 \times 10^{-24}$, while for the $Z-$spatial LIV coefficients $c_{e e}^{ZZ}-c_{\tau \tau}^{ZZ} < 5.1 \times 10^{-24}$ and $c_{\mu \mu}^{ZZ}-c_{\tau \tau}^{ZZ} < 7.1 \times 10^{-24}$, all of them at 95\% confidence level.
For instance, the null observation of LIV at the IceCube neutrino observatory constraint Re$(c_{ee})<6\times10^{-33}$ and Re$(c_{\tau \tau})< 8\times 10^{-34}$ with a Bayes factor $>31.6$~\cite{IceCube:2021tdn}.
\begin{figure}[H]
\begin{subfigure}[h]{0.49\textwidth}
			\caption{  }
			\label{f3a}
\includegraphics[width=\textwidth]{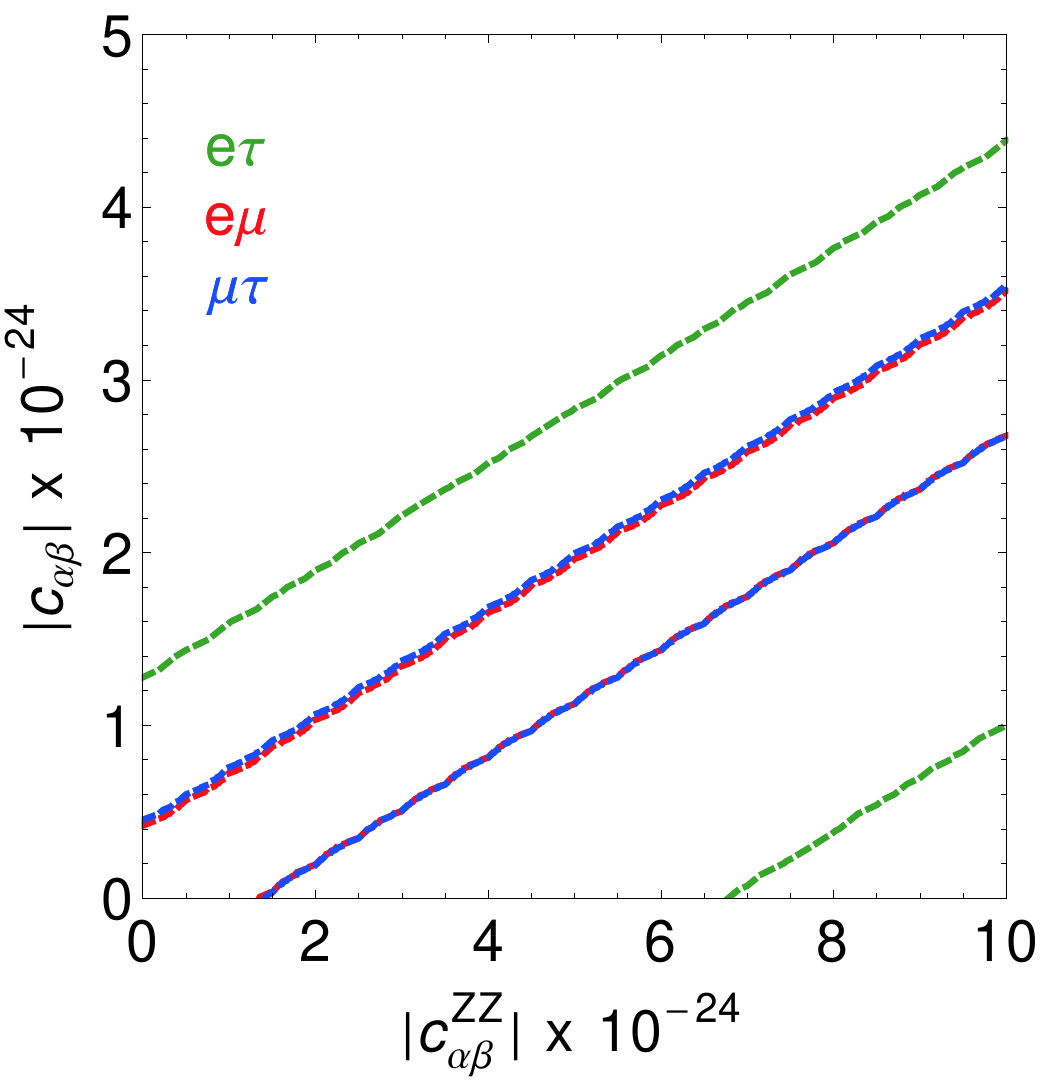}
		\end{subfigure}
		\hfill
		\begin{subfigure}[h]{0.495\textwidth}
			\caption{}
			\label{f3b}
			\includegraphics[width=\textwidth]{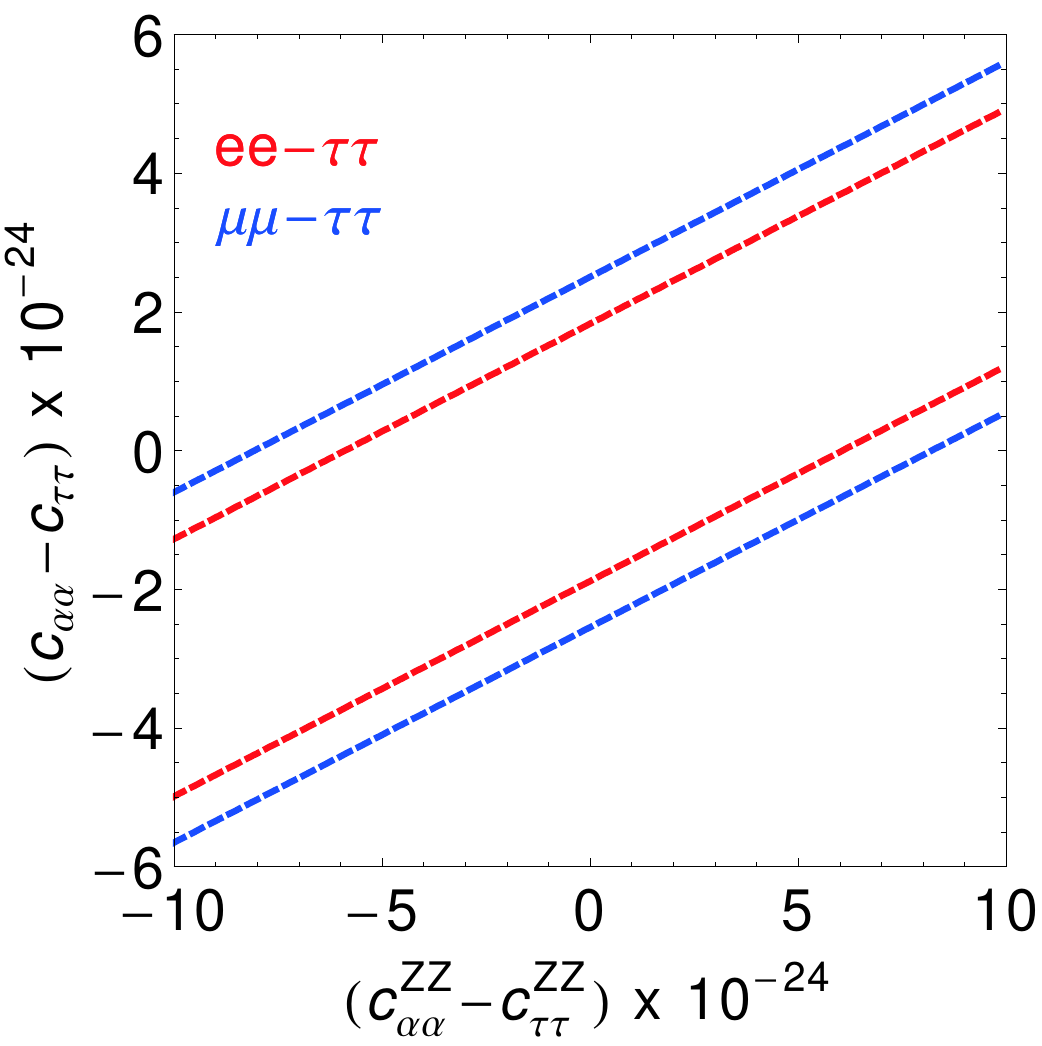}
		\end{subfigure}
		\hfill	
		 \caption{Projected 95\% C.L. sensitivities at the DUNE (TDR setup), in the $|c^{ZZ}_{\alpha\beta}|-|c_{\alpha\beta}|$ and $(c^{ZZ}_{\alpha\alpha}-c^{ZZ}_{\tau\tau})-(c_{\alpha\alpha}-c_{\tau \tau})$ planes, respectively. Here, we have marginalized over the corresponding LIV phases $\phi_{\alpha \beta}$ and $\phi_{\alpha \beta}^{ZZ}$ from [0$-2\pi$], as well as $\theta_{23}$ and $\delta_{CP}$, considering a 1$\sigma$ uncertainty of 10\% and 15\%, respectively. All the remaining oscillation parameters were fixed to their NO best fit values~\cite{deSalas:2020pgw}. Refer to the text for details.}
  \label{f3dune}
\end{figure}

In Fig.~\ref{f3dune}, we display the expected 95\% C.L. sensitivity regions for the case of the DUNE setup in the $|c_{\alpha \beta}^{ZZ}|-|c_{\alpha \beta}|$ and $(c_{\alpha \alpha}^{ZZ}-c_{\tau \tau}^{ZZ})-(c_{\alpha \alpha}-c_{\tau \tau} )$ planes, left and right panel, respectively. We have marginalized over the corresponding LIV phases $\phi_{\alpha \beta}$ and $\phi_{\alpha \beta}^{ZZ}$ from [0$-2\pi$], as well as the atmospheric mixing angle $\theta_{23}$ and the leptonic $CP$ phase $\delta_{CP}$, considering a 1$\sigma$ uncertainty of 10\% and 15\%~around their NO best-fit values, as displayed in Table~\ref{tab:1}. It is observed that flavor-changing LIV coefficients from the ($e-\mu$) and ($\mu-\tau$) sectors may be constrained approximately three times more effectively by the DUNE configuration than by those from the ($e-\tau$) sector, namely $|c_{e \tau}| <1.3 \times 10^{-24}~(|c_{e \tau}^{ZZ}|=0)$, and $|c_{e \tau}^{ZZ}| <6.9 \times 10^{-24}~(|c_{e \tau}|=0)$, all limits at 95\% C.L.

\begin{figure}[H]
\begin{subfigure}[h]{0.49\textwidth}
			\caption{  }
			\label{f4a}
\includegraphics[width=\textwidth]{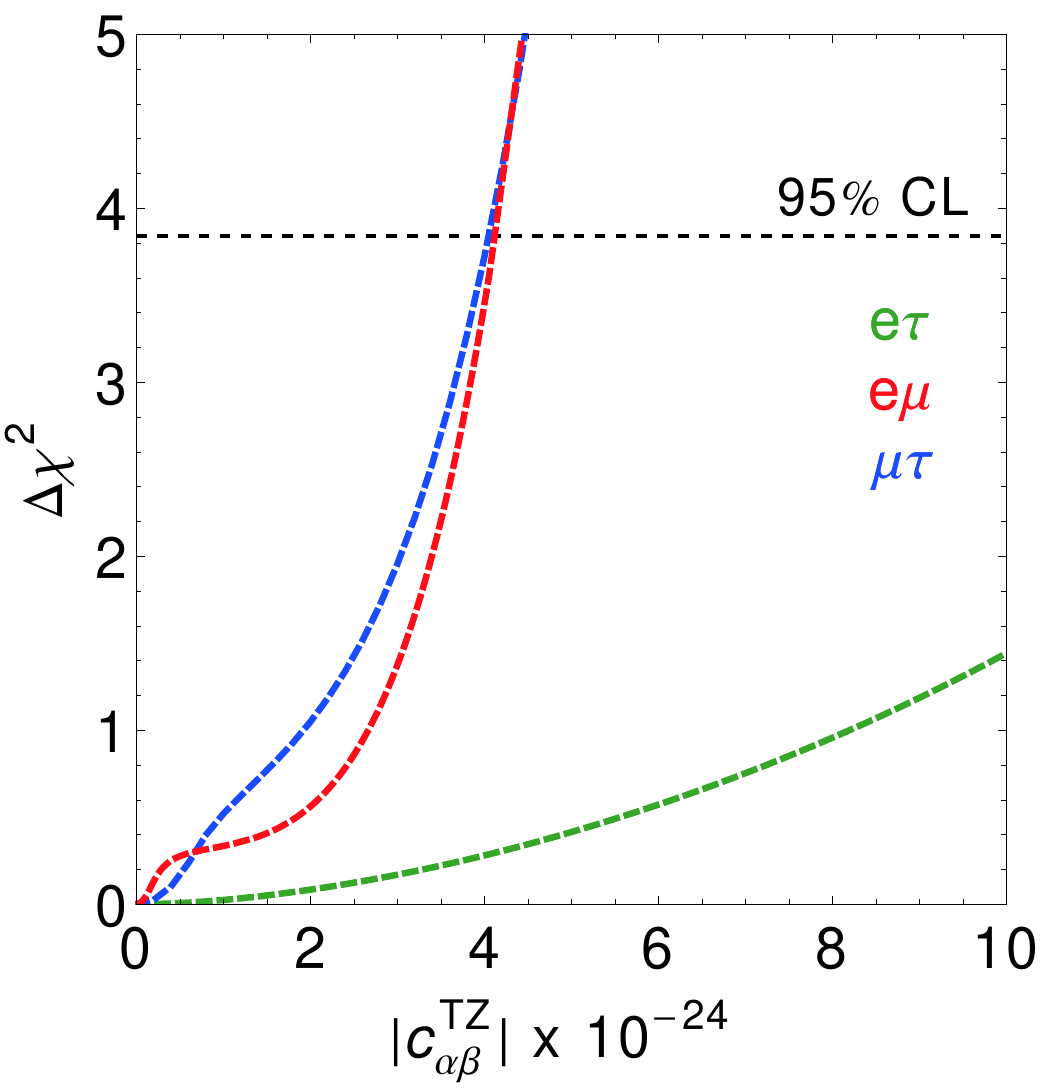}
		\end{subfigure}
		\hfill
		\begin{subfigure}[h]{0.49\textwidth}
			\caption{}
			\label{f4b}
			\includegraphics[width=\textwidth]{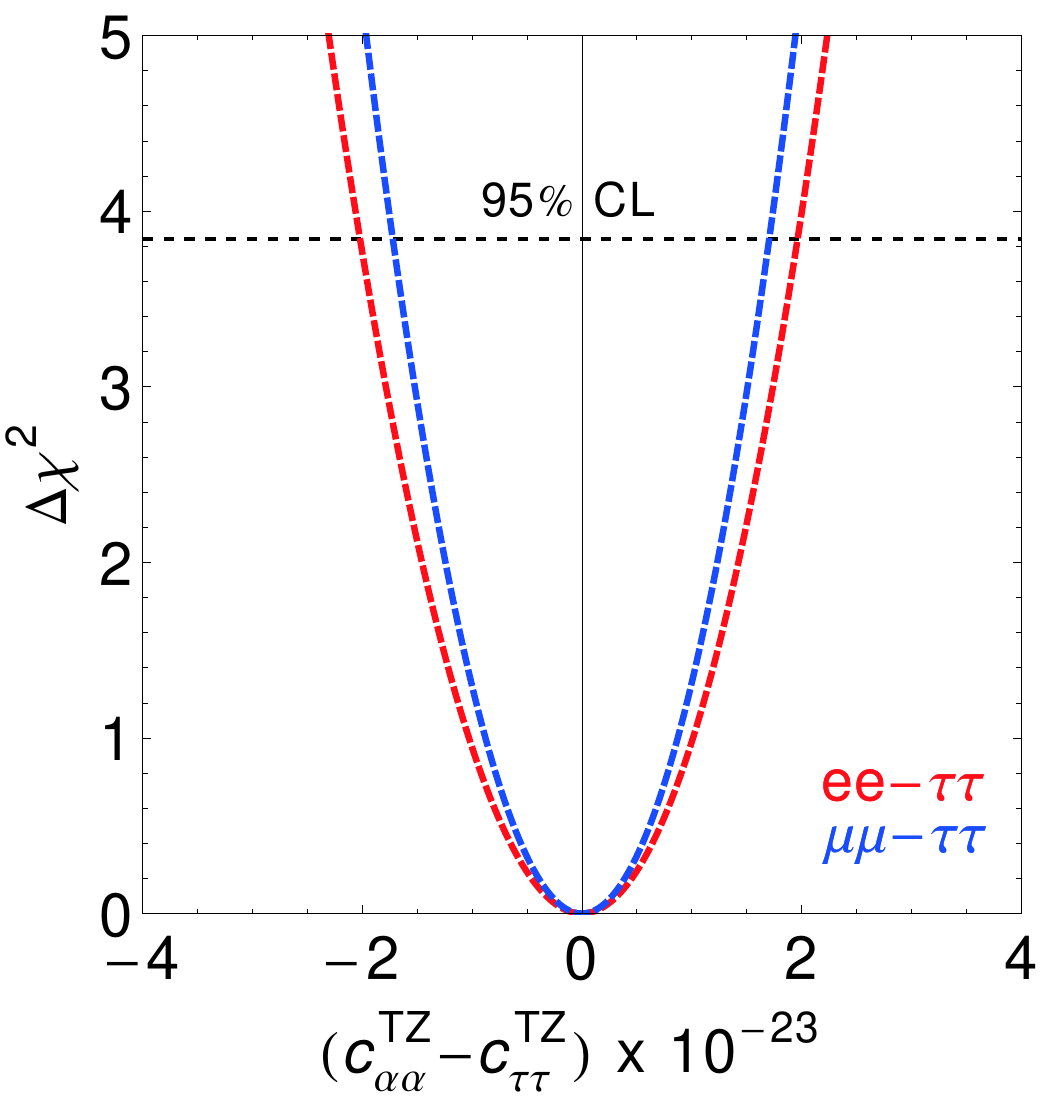}
		\end{subfigure}
		\hfill	
		 \caption{Expected 95\% C.L. sensitivities $|c^{TZ}_{\alpha\beta}|$ and $c^{TZ}_{\alpha\alpha}-c^{TZ}_{\tau\tau}$ at the DUNE (TDR setup). Here, we have marginalized over the corresponding LIV phases $\phi_{\alpha \beta}^{TZ}$ from [0$-2\pi$], as well as $\theta_{23}$ and $\delta_{CP}$, considering a 1$\sigma$ uncertainty of 10\% and 15\%, respectively. All the remaining oscillation parameters were fixed to their NO best fit values~\cite{deSalas:2020pgw}. Refer to the text for details.}
  \label{f4dune}
\end{figure}
In Fig.~\ref{f4dune}, we show in left (right) panel, the projected 95\% C.L. sensitivities to the SME coefficients $|c^{TZ}_{\alpha\beta}|$ and $c^{TZ}_{\alpha\alpha}-c^{TZ}_{\tau\tau}$, considering the DUNE configuration. We have marginalized over the corresponding LIV phases $\phi_{\alpha \beta}^{TZ}$ from [0$-2\pi$], as well as the atmospheric mixing angle $\theta_{23}$ and the leptonic $CP$ phase $\delta_{CP}$, considering a 1$\sigma$ uncertainty of 10\% and 15\%, accordingly. We observe that the sensitivities in the $TZ-$sector $(|c^{TZ}_{e\mu}|,~|c^{TZ}_{\mu \tau}|)\lesssim 4.2 \times10^{-24}$, are not as competitive as those from the isotropic and $Z-$spatial sectors $(|c_{e \mu}|,~|c_{\mu \tau}|)\lesssim 0.35 \times 10^{-24}$, and $(|c_{e \mu}^{ZZ}|,~|c_{ \mu \tau}^{ZZ}|) \lesssim 1.2 \times 10^{-24}$, as illustrated in the left and right panels of Fig.~\ref{f1dune}. Which can be partially understood from their functional dependence in the Hamiltonian $H_{\text{LIV}}$ of LIV (Eq.~\ref{LIVHAM})
\begin{equation}
    - H_{\text{LIV}} \simeq \frac{E_\nu}{2} \big[ 3 c_{\alpha \beta} -c_{\alpha \beta}^{ZZ} -2\hat{N}_Z c_{\alpha \beta}^{TZ} +\mathcal{O}(\hat{N}_Z^2) \big], ~~ \hat{N}_Z \sim 0.1~.
\end{equation}
Besides, the null observation of Lorentz violation employing the low-energy excess data set of the MiniBooNE experiment set $c^{TZ}_{e \mu} < (5.9 \pm 1.7) \times10^{-20}$~\cite{Katori:2010nf}, while bounds from the Double Chooz experiment constraint Re$(c^{TZ}_{e \tau})< 3.2 \times 10^{-17}$~\cite{Katori:2013jca, Kostelecky:2008ts}.

\begin{table}[H]
\caption{\label{tab:2} Current limits and projected sensitivities (shown in parenthesis), last column shows the interaction energy scale $M_*$ associated to the SME coefficients $c_{\alpha \beta}^{\mu \nu}$.}
\centering
\begin{tabular}{ c  c c }
\hline \hline
~LIV sector~ & Limit (Sensitivity) & Interaction energy scale \\
\hline 
~ neutrino~~&$\text{Re}(c_{\mu \tau}) < 7 \times 10^{-34}$, IceCube~\cite{IceCube:2021tdn}~~& $M_* \gtrsim 10^7~\text{eV}~\big( \rho_{\varphi, \odot}^\text{DM}\big)$~\\
~ neutrino~~&$|c_{e \mu}| < 8.0 \times 10^{-27}$, Super-Kamiokande~\cite{Super-Kamiokande:2014exs}~~& $M_* \gtrsim 10^5~\text{eV}~\big( \rho_{\varphi, \odot}^\text{DM}\big)$~\\
~ neutrino~~&$|c_{e \tau}| < 9.3 \times 10^{-25}$, Super-Kamiokande~\cite{Super-Kamiokande:2014exs}~~& $M_* \gtrsim 3 \times 10^4~\text{eV}~\big( \rho_{\varphi, \odot}^\text{DM}\big)$~\\
~ neutrino~~& $c_{e \mu}^{ZZ}< (2.6 \pm 0.8)\times 10^{-19}$, MiniBooNE~\cite{Katori:2010nf, Kostelecky:2008ts} & $M_* \gtrsim 2 \times 10^2~\text{eV}~\big( \rho_{\varphi}^\text{DE}\big)$~\\
~ neutrino~~& Re$(c^{ZZ}_{e \tau})< 4.9 \times 10^{-17}$, Double Chooz~\cite{Katori:2013jca} & $M_* \gtrsim 6 \times 10^1~\text{eV}~\big( \rho_{\varphi}^\text{DE}\big)$~\\
~ neutrino~~&($|c_{\alpha \beta}| \sim [1-10]\times 10^{-25}$),~DUNE~\cite{ Raikwal:2023lzk, Agarwalla:2023wft} &  $M_* \sim [3-6]\times 10^4~\text{eV}~\big( \rho_{\varphi, \odot}^{\text{DM}} \big)$~\\
~ neutrino~~&($|c_{\alpha \beta}| \sim [1-10]\times 10^{-25}$),~DUNE (this work) &  $M_* \sim [3-6]\times 10^4~\text{eV}~\big( \rho_{\varphi, \odot}^{\text{DM}} \big)$~\\ 
~ neutrino~~&($|c_{\alpha \beta}^{ZZ}| \sim [1-10]\times 10^{-24}$),~DUNE (this work) &  $M_* \sim [2-3]\times 10^3~\text{eV}~\big( \rho_{\varphi}^{\text{DE}} \big)$~\\
~ neutrino~~&Re$(c_{ee})<6\times10^{-33}$, IceCube~\cite{IceCube:2021tdn} & $M_* \gtrsim 6 \times 10^6~\text{eV}~\big( \rho_{\varphi, \odot}^\text{DM}\big)$~\\
~ neutrino~~&Re$(c_{\tau \tau})< 8\times 10^{-34}$, IceCube~\cite{IceCube:2021tdn} & $M_* \gtrsim 10^7~\text{eV}~\big( \rho_{\varphi, \odot}^\text{DM}\big)$~\\
~ neutrino~~&($c_{e e}-c_{\tau \tau} \simeq 1.3 \times 10^{-24}$),~DUNE (this work) &  $M_* \sim 3\times 10^4~\text{eV}~\big( \rho_{\varphi, \odot}^{\text{DM}} \big)$~\\ 
~ neutrino~~&($c_{\mu \mu}-c_{\tau \tau} \simeq 2.2 \times 10^{-24}$),~DUNE (this work) &  $M_* \sim 3\times 10^4~\text{eV}~\big( \rho_{\varphi, \odot}^{\text{DM}} \big)$~\\
~ neutrino~~&~~($c_{e e}^{ZZ}-c_{\tau \tau}^{ZZ} \simeq 5.0 \times 10^{-24}$),~DUNE (this work)~&  $M_* \sim 2\times 10^3~\text{eV}~\big( \rho_{\varphi}^{\text{DE}} \big)$~ \\ 
~ neutrino~~&~~($c_{\mu \mu}^{ZZ}-c_{\tau \tau}^{ZZ} \simeq 7.0 \times 10^{-24}$),~DUNE (this work)~&  $M_* \sim 2\times 10^3~\text{eV}~\big( \rho_{\varphi}^{\text{DE}} \big)$~\\
~ electron~~&$-80\lesssim c_{ee}/ 10^{-20}\lesssim 4$~\cite{Kostelecky:2008ts} \\
 \hline \hline
\end{tabular}
\end{table}

\section{Conclusions}
\label{conclusion}
In contemporary cosmology, one of the most fascinating puzzles are the dark matter and dark energy conundrum. One of the most common fields employed to describe dark energy is the scalar field. Furthermore, a scalar field can also characterize ultralight dark matter, which is one of the most promising alternatives for dark matter. 

On the other hand, if the $CPT$ and Lorentz symmetries are broken at very high energies, well beyond the electroweak scale, oscillations of high energy neutrinos may probe energy scales where this possible violations arise. The aforementioned violation of Lorentz invariance might arise from neutrino non-standard interactions with scalar fields.

In this paper, we have outlined the case of an effective tensorial neutrino--scalar field interaction. The corresponding scalar field could be identified as either a dark energy or dark matter candidate. Moreover, within this framework, the effective neutrino interaction with cosmological scalar fields can be associated to the $CPT-$even SME coefficients via the energy momentum-tensor $T^{\mu \nu}_{\varphi}$ (Sec.~\ref{model}). In the case of the isotropic coefficients $c_{\alpha \beta}$, a simple relation can be established in terms of the energy densities ($\rho_{\varphi}$) for DE or DM (Eq.~\ref{isoc}). Therefore, bounds and sensitivities set on the $CPT-$even SME coefficients ($c^{\mu \nu}_{\alpha \beta}$) may be related to the energy scale ($M_*$) of this interaction.

Furthermore, we estimate sensitivities to the dimension-four $CPT-$even SME coefficients $c^{\mu \nu}_{\alpha \beta}$ of the SME, particularly the isotropic $c_{\alpha \beta}$ and $Z-$spatial $c_{\alpha \beta}^{ZZ}$. 
For an accelerator-based experiment similar to DUNE, our predicted sensitivities at 95\% C.L. are $|c_{\alpha \beta}| \sim [1-10]\times 10^{-25}$ (left panel of Fig.~\ref{f1dune}) and  $|c_{\alpha \beta}^{ZZ}| \sim [1-10]\times 10^{-24}$ (right panel of Fig.~\ref{f1dune}), while for the diagonal LIV coefficients $c_{e e}-c_{\tau \tau} \simeq 1.3 \times 10^{-24}$ and $c_{\mu \mu}-c_{\tau \tau} \simeq 2.2 \times 10^{-24}$ (left panel of Fig.~\ref{f2dune}); $c_{e e}^{ZZ}-c_{\tau \tau}^{ZZ} \simeq 5.0 \times 10^{-24}$ and $c_{\mu \mu}^{ZZ}-c_{\tau \tau}^{ZZ} \simeq 7.0 \times 10^{-24}$ (right panel of Fig.~\ref{f2dune}). 
Our results are summarized in Table~\ref{tab:2}.

Upcoming and present neutrino experiments such as the DUNE, KM3NeT, IceCube-Gen2 and GRAND proposals; as well as the IceCube neutrino observatory, could shed more light on these types of neutrino interactions.

This paper represents the views of the authors and should not be considered a DUNE collaboration paper.

\section*{Acknowledgments}
This work was partially supported by SNII-M\'exico and CONAHCyT research
Grant No.~A1-S-23238. Additionally the work of R.~C. was partially supported
by COFAA-IPN, Estímulos al Desempeño de los Investigadores (EDI)-IPN and SIP-IPN Grant No.~20241624.

\appendix

\section{Details of a tensorial neutrino interaction}
\label{appx}

A possible realization of the tensorial coupling discussed in this work could arise from conformal coupling and disformal transformations where the scalar field couples to different kinds of matter fields in the form \cite{Gauthier:2009wc,Brax:2012hm,Brax:2014vva,CarrilloGonzalez:2020oac,YazdaniAhmadabadi:2022low,YazdaniAhmadabadi:2022wiq}
\begin{equation}
    S_M = \int d^ 4 x \sqrt{-\tilde{g}} {\mathcal{L}}_m (\tilde{\Psi}_i, \tilde{g}^i _{\mu \nu}) \,\, ,
\end{equation}
where $\tilde{g}$ is the determinant of the metric $\tilde{g}^i _{\mu \nu}$ and the disformal transformation can be written as \cite{Bekenstein:1992pj,Deffayet:2020ypa,Zumalacarregui:2013pma}
\begin{equation}
    \tilde{g}_{\mu \nu} = a(\varphi, X)g_{\mu \nu} + b(\varphi, X) \partial_\mu \varphi \partial_\nu \varphi \,\, .
\end{equation}
Disformal transformations have been applied to a wide range of topics in cosmology, from inflation \cite{Kaloper:2003yf} through dark matter \cite{Bekenstein:2004ne}, dark energy \cite{Zumalacarregui:2010wj} and its cosmological implications \cite{vandeBruck:2013yxa}. Besides, disformal transformations have been applied in generalized Palatini gravities~\cite{Olmo:2009xy}, and it has been studied their possible non-trivial effects on radiation and signatures in laboratory tests~\cite{Philippe:Brax_2012}.
Following the construction presented in \cite{Brax:2012hm,Brax:2014vva} we can decompose the transformation in the following form \footnote{In fact, there are more general disformal transformations akin~\cite{Ikeda:2023ntu}, $\bar{g}_{\mu \nu} = F_0 g_{\mu \nu} + F_1 \varphi_\mu \varphi_\nu + 2F_2\varphi_{(\mu} X_{\nu )} + F_3 X_\mu X_\nu$ where $\varphi_\mu := \nabla_\mu \varphi$, $X_\mu := \nabla _\mu X$. The functions $F_i$ depend on the variables $\varphi, X, Y, Z$, where $Y:=\varphi^\mu X_\mu$ and $Z:= X^\mu X_\mu$. However, the consistency of the fermionic coupling requires that $F_3 =0$ \cite{Takahashi:2022ctx}. }
\begin{equation}
   \tilde{g}_{\mu \nu} = g_{\mu \nu} + \alpha(\varphi, X, M_*)g_{\mu \nu} + \frac{\beta(\varphi, X)}{M_* ^4} \partial_\mu \varphi \partial_\nu \varphi = g_{\mu \nu} + h_{\mu \nu}  \,\, , 
\end{equation}
where $M_*$ is related to energy scale of the interaction. Considering the $h_{\mu \nu}$ term as a small correction to $g_{\mu \nu}$, we can expand the action at first order and obtain a derivative coupling of the scalar field with matter 
\begin{equation}
S_M = \int d^ 4 x \sqrt{-{g}} {\mathcal{L}}_m ({\Psi}_i, {g}^i _{\mu \nu}) +\sum_i \int d^ 4 x \sqrt{-{g}}\frac{1}{2} T^{\mu \nu}_i \left(\alpha(\varphi, X, M_*)g_{\mu \nu} + \frac{\beta(\varphi, X)}{M_*^4} \partial_\mu \varphi \partial_\nu \varphi\right) \,\, ,    
\end{equation}
where the sum considers different matter components and $T^{\mu \nu} =\frac{2}{\sqrt{-g}}\frac{\delta S_M}{\delta g_{\mu \nu}}$ is the energy momentum tensor of the matter fields. In the case of fermions 
\begin{equation}
    S_M = \int d^ 4 x \sqrt{-{g}}\frac{-i}{2}(\bar{\Psi} \gamma^\mu D_\mu \Psi -D_\mu \bar{\Psi}\gamma^\mu \Psi  + 2im \bar{\Psi} \Psi)\, , 
\end{equation}
and the energy-momentum tensor is
\begin{equation}
   T_{\mu \nu}^{\Psi} = -\frac{i}{2}(\bar{\Psi} \gamma_{(\mu} D_{\nu )} \Psi -D_{(\mu} \bar{\Psi}\gamma_{\nu )} \Psi) \,.   
\end{equation}
Therefore, the interaction term can be written as
\begin{equation}
   S_I = -\int d^ 4 x \sqrt{g} \frac{i}{4}(\bar{\Psi} \gamma_{(\mu} D_{\nu )} \Psi -D_{(\mu} \bar{\Psi}\gamma_{\nu )} \Psi )\left(\alpha(\varphi, X,M_*)g^{\mu \nu} + \frac{\beta(\varphi, X)}{M_* ^4}\partial^\mu \varphi \partial^\nu \varphi\right) \,. 
\end{equation}
It is convenient to select a particular choice of the $\alpha$ and $\beta$ functions to construct the energy-momentum tensor $T_\varphi ^{\mu \nu}$ for the scalar field $\varphi$ in the interaction
\begin{equation}
    S_I = \int d^ 4 x \sqrt{-g} \frac{i k}{2M_*^4 }(\bar{\Psi} \gamma_{(\mu} D_{\nu )} \Psi + \text{h.c.}\,)T_\varphi ^{\mu \nu} \, .
\end{equation}
The effective Lagrangian has a similar structure that appears in modified gravity~\cite{Asimakis:2022jel} where there is a coupling of the energy-momentum tensor with the Einstein tensor and a proposal of a coupling among the energy-momentum tensor and vector and scalar fields \cite{BeltranJimenez:2018tfy}. Moreover, a similar interaction was proposed in Refs.~\cite{Kehagias:2011cb, Ciuffoli:2011ji} in the context of Galileon$-$neutrino couplings. For neutrinos, the interaction term can be written as
\begin{equation}
    S_I = \int d^ 4 x \sqrt{-g} \frac{ i \lambda_{\alpha \beta}}{2 M_*^4 }(\bar{\nu}_\alpha \gamma_{(\mu}(1-\gamma_5) D_{\nu )} \nu_\beta + \text{h.c.}\,)T_\varphi ^{\mu \nu} \, .
\end{equation}
In cases where the Minkowski metric ($\eta_{\mu \nu}$) is a good approximation, the former relation reduces to
\begin{equation}
  S_I = \int d^ 4 x \sqrt{-\eta} \frac{ i \lambda_{\alpha \beta}}{2 M_*^4 }(\bar{\nu}_\alpha \gamma_{\mu}(1-\gamma_5) \partial_{\nu } \nu_\beta + \text{h.c.}\,)T_\varphi ^{\mu \nu} \, .  
\end{equation}

\end{document}